\documentclass[twocolumn,showpacs,amsmath,amssymb]{revtex4}
\input{epsf}

\usepackage{graphicx}
\usepackage{array}
\usepackage{longtable}
\usepackage{rotating,booktabs}
\usepackage{booktabs,threeparttable}
\usepackage{bm}

\begin{document}

\title{Hyperfine effects on potassium tune-out wavelengths and polarizabilities}

\author{Jun Jiang and J. Mitroy}

\affiliation{ School of Engineering, Charles Darwin
University, Darwin NT 0909, Australia}

\date{\today}

\begin{abstract}

The influence of hyperfine interactions on the tune-out wavelengths of
the $^{39,40,41}$K isotopes of the potassium atom was investigated.  The
hyperfine interaction of the $4s_{1/2}$ ground state results in a shift and splitting
of the primary tune-out wavelength near 769 nm.  The $4s_{1/2}$ state
hyperfine splittings of the primary tune-out wavelength were almost equal to the
hyperfine splittings of the ground states.  The splittings in the wavelengths
were 0.0008, 0.0027 and 0.0005 nm for $^{39}$K, $^{40}$K and $^{41}$K respectively.
The hyperfine splitting of the
$np_{J}$ levels leads to the creation of additional tune-out
wavelengths.  The additional tune-out wavelengths could be difficult to
detect due to very small differences from the transition wavelengths to the
$4p_{J,F}$ states. The hyperfine Stark
shift for the ground states of all three isotopes were also computed
and the value for $^{30}$K was found to be compatible with the previous
experiments and the most recent calculation using relativistic many body
perturbation theory.
\end{abstract}

 \maketitle

\section{Introduction}

The primary tune-out wavelength for the $4s_{1/2}$ ground state of neutral
potassium located between the $4p_{1/2}$ and $4p_{3/2}$ excitations
has recently measured to be 768.9712(15) nm \cite{holmgren12a}.
There have also been calculations of a number of tune-out wavelengths
for potassium \cite{arora11a,jiang13a}.  The present
work extends those calculations to take into consideration the impact
of hyperfine interactions on the tune-out wavelengths for the
$^{39,40,41}$K isotopes.  Plans to measure tune-out wavelengths to
increased precision mean that hyperfine splitting could easily become
significant at the level of anticipated precision \cite{holmgren12a,cronin13a}.
The present article determines the impact of the hyperfine structure
upon the tune-out wavelengths of potassium.  We use the matrix elements
previously used to determine the tune-out wavelengths of potassium
\cite{jiang13a} and minimal details of this previous calculation are
presented.

The concept of the tune-out wavelength was initially introduced as
a means to release one atom species from a two species optical lattice
\cite{leblanc07a}.  Other applications of tune-out wavelengths within
two species optical lattices \cite{arora11a} have recently been discussed.
One of possible application would be the measurements of tune-out
wavelengths to make precise estimates of oscillator strength
ratios \cite{arora11a}.  A relative precision of about 0.2$\%$ has been achieved
in the $5s \to 6p_J$ transition matrix elements of rubidium
\cite{herold12a} and a relative precision of
0.19$\%$ has been achieved for the $(4s \to 4p_{1/2}):(4s \to 4p_{3/2})$
line strength ratio
for potassium \cite{holmgren12a}.  There are relatively few atomic
oscillator strengths measured at this level of precision
\cite{bouloufa09a}.

The present investigation entails the determination of the polarizability
of the K($4s$) ground state for different hyperfine levels.  The calculations
are similar to those used in the determination of the hyperfine Stark shift
and there have been a number of experiments and calculations on hyperfine
Stark shifts for the alkali atoms
\cite{mitroy10a,simon98a,johnson08a,safronova08b,godone05a,beloy06a,angstmann06a}.
The most important of these investigations
are those related to the determination of the blackbody radiation
shift of the $^{137}$Cs clock transition \cite{angstmann06a}.

The approach used to determine the influence of the hyperfine shift upon
the dynamic polarizability is somewhat unorthodox.  Most calculations of
the hyperfine Stark shift are performed using third order perturbation
theory \cite{kelly73a,lee75a,beloy06a,angstmann06a}.  The
hyperfine interaction operator alters the polarizability in two
distinctly different ways.  First of all, the
hyperfine energy shifts directly enter the denominator in the
oscillator strength sum rule used to calculate the dipole polarizability.
Second, the hyperfine interactions leads to the mixing of the ground
$4s_{1/2}$ state with excited bound $ns_{1/2}$ and continuum
$\varepsilon_{1/2}$ states.  This results in a change of the
$4s_{1/2} \to 4p_{3/2}$ line strengths.   These two effects
have roughly same influence on the polarizability.

The approach used to calculate the hyperfine polarizabilities utilizes
existing information about the hyperfine interaction constants to
determine the hyperfine energy shifts of the low lying states of
potassium.  We do not use a perturbation theory calculation to
determine the impact of the hyperfine interaction on the line
strength.  Instead, the dipole matrix element from the resonant
transition is treated as a parametric function of the binding energy,
and matrix elements adjusted for the hyperfine energy shifts are
used in the calculation of the polarizability.  The method was
validated by a comparison with the previous calculated and
experimental hyperfine Stark shifts for
$^{39}$K \cite{safronova08b,kelly73a,lee75a,snider66a,mowat72a}.
Shifts in the primary tune-out wavelengths for the  ground states of
$^{39,40,41}$K were found to be very closely related
to the hyperfine energy shifts of the ground states.

\section{Formulation and calculations}

\subsection{Hyperfine splitting }

The first part of our calculation required the determination of the hyperfine energy
shifts of the ground and low-lying excited states.   Rather than calculate this
information directly, the information is sourced from experiment or other
calculations.  The potassium dipole polarizability is dominated by the resonant
transition, so hyperfine splitting is only taken into consideration for the
ground state and some of the low-lying excited states.

According to first-order perturbation theory, the energy for a hyperfine
state $|LJIF\rangle $ is given \cite{arimondo77a,armstrong71a} by
\begin{eqnarray}
E = E_{NLJ} + W_F
\end{eqnarray}
where $E_{NLJ}$ is the energy for the unperturbed fine structure state and $W_F$ is the
hyperfine interaction energy.  The hyperfine interaction energy can be written as
\begin{eqnarray}
W_F = \frac{1}{2}AR+B\frac{\frac{3}{2}R(R+1)-2I(I+1)J(J+1)}{2I(2I-1)2J(2J-1)}
\end{eqnarray}
where,
\begin{eqnarray}
R=F(F+1)-I(I+1)-J(J+1),
\end{eqnarray}
$F$ is total angular momentum, $I$ is the nuclear spin, $J$ is the total electronic
angular momentum of the associated fine structure state, and $A$ and $B$ are hyperfine
structure constants.  It is usual to give the $A$ and $B$ coefficients in MHz where
$1.0 \ {\rm MHz} = 1.519829903\times 10^{-10}$ a.u.  The energies of the different
hyperfine states of $^{39}$K,  $^{40}$K and $^{41}$K are listed in Table \ref{tab1}.
The $^{39}$K and $^{41}$K isotopes are stable, while $^{40}$K has a very long
lifetime.   The $^{39}$K isotope is the most common at 93.3$\%$ abundance.
These tabulations are only include the $4s_{1/2}$, $4p_J$ and $5p_J$ states.
The hyperfine energy shifts are largest for $^{40}$K. The energy shifts for the
$4s_{1/2}$ ground states are about an order of magnitude larger than those of
the $4p_{1/2}$ excited states.  Similarly, the hyperfine splitting for the
$np_{1/2}$ states are significantly larger than the splitting of the
$np_{3/2}$ states.

\begin{table*}
\caption{\label{tab1} The binding energies of the hyperfine states of $^{39}$K,
$^{40}$K and $^{41}$K relative to the K$^{+}$ core.
The nuclear spin, $I$ is indicated.  The notation $a[b]$ means $a \times 10^{b}$.
The energies of the different isotopes are normalized so that the energies of the
$5p_J$ levels are the same.  Hyperfine interaction constants are sourced from
experiment with two exceptions which are denoted by labelling
with (T).  The absolute binding energies are known to eight
significant digits so the digits beyond that should only be interpreted as having
significance when comparing the energy differences of the hyperfine
levels of the same isotope.
}
\begin{ruledtabular}
\begin {tabular}{lcccccc}
                & $E_{nLJ}$ (a.u.)  & $A$ (MHz)                           &    $B$  (MHz)               &  $F$    &  $W_F$ (a.u.) &  $E_{nLJIF}$ (a.u.) \\
\hline
 \multicolumn{7}{c}{$^{39}$K, $I = 1/2$ }  \\
      $4s_{1/2}$ & $-$0.15951645       &   230.859860 \cite{arimondo77a}   &    -                  &  1    &$-$4.3858[$-$8]   &  $-$0.159516493858 \\
                 &                     &                                  &                        &  2    &  2.6315[$-$8]   &  $-$0.159516423685 \\
      $4p_{1/2}$ & $-$0.10035159       &  27.775 \cite{falke06b}          &-                       &  1    &$-$5.2767[$-$9]   &  $-$0.100351595277 \\
                 &                     &                                  &                        &  2    &  3.1660[$-$9]   &  $-$0.100351586834 \\
      $4p_{3/2}$ & $-$0.100088643      &  6.093 \cite{falke06b}           &2.786 \cite{falke06b}   &  0    &$-$2.9433[$-$9]   &  $-$0.100088645943 \\
                 &                     &                                  &                        &  1    &$-$2.4407[$-$9]   &  $-$0.100088645441 \\
                 &                     &                                  &                        &  2    &$-$1.0121[$-$9]   &  $-$0.100088644012 \\
                 &                     &                                  &                        &  3    &  2.1894[$-$9]   &  $-$0.100088640811 \\
      $5p_{1/2}$ & $-$0.0469686695     &$-$9.02 \cite{arimondo77a}         &    -                    &  1    &   $-$1.7136[$-$9]      &  $-$0.0469686712136 \\
                 &                     &                                  &                        &  2     &   1.0281[$-$9]      &  $-$0.0469686684718 \\
      $5p_{3/2}$ & $-$0.0468832095     &$-$1.969 \cite{arimondo77a}       &$-$0.87 \cite{arimondo77a}  &  0    &  $-$9.5692[$-$10]      &  $-$0.0468832104569 \\
                 &                     &                                  &                        &  1    &   $-$7.8989[$-$10]      &  $-$0.0468832102899 \\
                 &                     &                                  &                        &  2    &   $-$3.2361[$-$10]     &  $-$0.0468832098236 \\
                 &                     &                                  &                        &  3    &     7.0638[$-$10]      &  $-$0.0468832087936 \\
 \multicolumn{7}{c}{$^{40}$K, $I = 4$  }  \\
      $4s_{1/2}$ & $-$0.15951648572    &$-$285.7308 \cite{arimondo77a}   &    -                       &  7/2    &      1.0857[$-$7]      &  $-$0.159516377150\\
                 &                     &                                  &                            &  9/2    &   $-$8.6852[$-$8]      &  $-$0.159516572568 \\
      $4p_{1/2}$ & $-$0.100351606621   &$-$34.52300 \cite{falke06b}      &    -                       &  7/2    &      1.3117[$-$8]      &  $-$0.100351593504 \\
                 &                     &                                  &                            &  9/2    &   $-$1.0494[$-$8]      &  $-$0.100351617115 \\
      $4p_{3/2}$ & $-$0.10008865956    &$-$7.585 \cite{falke06b}         &$-$3.445 \cite{falke06b}   &  5/2    &      8.3888[$-$9]      &  $-$0.100088651173 \\
                 &                     &                                  &                            &  7/2    &      4.7140[$-$9]      &  $-$0.100088654848 \\
                 &                     &                                  &                            &  9/2    &   $-$3.4733[$-$10]     &  $-$0.100088659909 \\
                 &                     &                                  &                            & 11/2    &   $-$7.0476[$-$9]      &  $-$0.100088666610 \\
      $5p_{1/2}$ & $-$0.0469686695     &$-$10.98(T) \cite{singh12a}         &    -                       &  7/2    &      4.1719[$-$9]      &  $-$0.0469686653281 \\
                 &                     &                                  &                           &  9/2     &   $-$3.3375[$-$9]      &  $-$0.0469686728375 \\
      $5p_{3/2}$ & $-$0.0468832095     &$-$2.45 \cite{arimondo77a}       &$-$1.16 \cite{arimondo77a}    &  5/2    &      2.7061[$-$9]      &  $-$0.0468832067939 \\
                 &                     &                                  &                            &  7/2    &      1.5240[$-$9]      &  $-$0.0468832079759 \\
                 &                     &                                  &                            &  9/2    &   $-$1.0904[$-$10]     &  $-$0.0468832096090 \\
                 &                     &                                  &                            & 11/2    &   $-$2.2782[$-$9]      &  $-$0.0468832117782 \\
 \multicolumn{7}{c}{$^{41}$K, $I = 3/2$ }  \\
      $4s_{1/2}$ & $-$0.1595165190       &   127.0069352 \cite{arimondo77a} &    -                   &  1    & $-$2.4129[$-$8]      &  $-$0.159516543159 \\
                 &                     &                                  &                        &  2    &    1.4477[$-$8]      &  $-$0.159516504553 \\
      $4p_{1/2}$ & $-$0.1003516232       &  15.245 \cite{falke06b}          &-                       &  1    & $-$2.8962[$-$9]      &  $-$0.100351626136 \\
                 &                     &                                  &                        &  2    &    1.7377[$-$9]      &  $-$0.100351621502 \\
      $4p_{3/2}$ & $-$0.1000886761      &  3.362 \cite{falke06b}           &3.351 \cite{falke06b}   &  0    & $-$1.2795[$-$9]      &  $-$0.100088677420 \\
                 &                     &                                  &                        &  1    & $-$1.2778[$-$9]      &  $-$0.100088677418 \\
                 &                     &                                  &                        &  2    & $-$7.6520[$-$10]     &  $-$0.100088676905 \\
                 &                     &                                  &                        &  3    &    1.2770[$-$9]      &  $-$0.100088674863 \\
      $5p_{1/2}$ & $-$0.0469686695     &$-$4.84(T) \cite{singh12a}         &    -                    &  1    &     $-$9.1950[$-$10]      &  $-$0.0469686704195 \\
                 &                     &                                  &                        &  2     &   5.5169[$-$10]      &  $-$0.0469686689483 \\
      $5p_{3/2}$ & $-$0.0468832095     &$-$1.08 \cite{arimondo77a,ney69a}        &1.06 \cite{arimondo77a,ney69a}  &  0    &   $-$4.1415[$-$10]      &  $-$0.0468832099142 \\
                 &                     &                                  &                        &  1    &$-$4.1111[$-$10]      &  $-$0.0468832099111 \\
                 &                     &                                  &                        &  2    &   $-$2.4393[$-$10]     &  $-$0.0468832097439 \\
                 &                     &                                  &                        &  3    &   4.0959[$-$10]      &  $-$0.0468832090904 \\
\end{tabular}
\end{ruledtabular}
\end{table*}

\subsection{Reduced Matrix Elements}

The dipole matrix elements between the different hyperfine levels are
computed from the original $L J$ tabulation \cite{jiang13a} using the
Wigner-Eckart theorem.  The original calculations of the potassium
tune-out wavelengths were explicitly relativistic.   These matrix elements,
after making a small
energy dependent correction described later, were then used to compute
the polarizabilities and subsequently the tune-out wavelengths.

The transition matrix elements between the two hyperfine levels $ |n_1L_1J_1IF_1\rangle $ and
$|n_2L_2J_2IF_2\rangle$ can be written as

\begin{eqnarray}
&&\langle L_2J_2IF_2 \| r^kC^k(r)\| L_1J_1IF_1\rangle = (-1)^{I+J_2+F_1+k} \nonumber \\
            & \times & \hat{F}_1 \hat{F}_2
  \left\{
\begin{array}{ccc}
I & J_1 & F_1 \\
k & F_2 & J_2 \\
\end{array}
\right \}
\langle L_2J_2 \| r^kC^k(r) \| L_1J_1 \rangle ,
\end{eqnarray}
where $k=1$ for a dipole transition and
$\hat{F} = \sqrt{2F+1}$.

The absorption oscillator strength $f^{(k)}_{gi}$ for
a dipole transition from hyperfine level $g \to i$
is defined in the $F$-representation
as
\begin{equation}
f^{(k)}_{gi} = \frac{ 2|\langle L_iJ_iIF_i \| r^kC^k(r) \| L_gJ_gIF_g\rangle |^2 \varepsilon_{gi} } {(2k+1)(2F_g+1)} \ ,
\end{equation}
where $\varepsilon_{gi}$ is the excitation energy of the transition.

The matrix elements, $A_{ij}$,  are treated as parametric functions of their
binding energies.  The functional form adopted is
\begin{eqnarray}
A_{ij}(E_i,E_j) &\approx&  A_{ij}(E_{0,i},E_{0,j}) + \frac{\partial A_{ij}}{\partial E_i}(E_i-E_{0,i})
         \nonumber \\
	 &+& \frac{\partial A_{ij}}{\partial E_j}(E_i-E_{0,j}) \ ,
\end{eqnarray}
where $E_{0,i}$ and $E_{0,j}$ are the binding energies without any
hyperfine splitting.  The original $(L,J)$ matrix element set was
computed using a single-electron approach with a semi-empirical polarization
potential \cite{jiang13a}.  The partial derivatives were evaluated
by redoing the calculations with a slightly different polarization
potential and noting the change in the reduced matrix elements.
The original $4s \to 4p_{J}$ and $4s \to 5p_{J}$ matrix elements of
the $(L,J)$ calculation and the matrix element derivatives are listed
in Table \ref{tab2}.

\begin{table}
\caption{\label{tab2} The partial derivatives describing the variation of the
matrix elements with respect to the initial and final state binding
energies.  }
\begin{ruledtabular}
\begin {tabular}{lccc}
Transition  & $A(E_{0i},E_{0j}) $  & $\frac{\partial A}{\partial E_{4s}}$  & $\frac{\partial A}{\partial E_{j}} $  \\  \hline
  $4s_{1/2}$-$4p_{1/2}$ &   4.102991192     & 28.334       & $-$0.548       \\
  $4s_{1/2}$-$4p_{3/2}$ &   5.801566158      & 40.258       & $-$1.267   \\
  $4s_{1/2}$-$5p_{1/2}$ &   0.2633449165    & $-$17.058    & 138.399       \\
  $4s_{1/2}$-$5p_{3/2}$ &   0.3886341459    & $-$23.872    & 193.610     \\
\end{tabular}
\end{ruledtabular}
\end{table}

\subsection{Handling isotopic effects}

Calculations of polarizabilities and tune-out wavelengths have also been done
for the $^{40,41}$K isotopes.  The references binding energies given in Table
\ref{tab1} are those of $^{39}$K.  The energies attributed to the $^{40,41}$K
isotopes in Table \ref{tab1} were computed as follows.  The energies
for the K($5p$) states were used as a reference point and held fixed.   The overall
binding energies of the $^{40,41}$K($4s$) and $^{40,41}$K($4p_J$) states were then
adjusted to be more tightly bound by including the isotope shifts listed in
Table \ref{tab3}.  The binding energies given in Table \ref{tab1} for the
K($4s$) and K($4p_J$) states incorporate these shifts.  It is necessary to
include these shifts since tune-out wavelengths depend sensitively on
the energy spacing of the K($np_J$) levels from the ground state.

\begin{table}
\caption{\label{tab3} Isotope shifts, $\delta \nu$, of the $4s-4p_{1/2}$,
$4s-4p_{3/2}$ and $4s-5p$ transitions of potassium.  The shifts are for
the mean energies once the effects of hyperfine splitting have been
removed.  The positive shifts mean that the transition energy would be
be larger for $^{40,41}$K than for $^{39}$K.  The notation $a[b]$ means
$a \times 10^{b}$. }
\begin{ruledtabular}
\begin {tabular}{lcc}
    &  (MHz)  & (a.u.)      \\ \hline
  \multicolumn{3}{c}{IS($4s-4p_{1/2}$) \cite{falke06b}}       \\
 $\delta\nu$($^{40}$K--$^{39}$K)  & 125.64(10) & 1.9095(16)[$-$8] \\
 $\delta\nu$($^{41}$K--$^{39}$K)  & 235.49(9)  & 3.5790(14)[$-$8] \\
  \multicolumn{3}{c}{IS($4s-4p_{3/2}$) \cite{falke06b}}    \\
 $\delta\nu$($^{40}$K--$^{39}$K)  & 126.03(15) & 2.0674(23)[$-$8] \\
 $\delta\nu$($^{41}$K--$^{39}$K)  & 236.18(17) & 3.5895(25)[$-$8] \\
  \multicolumn{3}{c}{IS($4s-5p$) \cite{behrle11a}}     \\
 $\delta\nu$($^{40}$K--$^{39}$K)  & 235.0(20) & 3.572(36)[$-$8]  \\
 $\delta\nu$($^{41}$K--$^{39}$K)  & 454.2(8)  & 6.903(14)[$-$8] \\
\end{tabular}
\end{ruledtabular}
\end{table}

While the separation between the K($4s,4p_J,5p_{J'})$ levels is
correct for the different isotopes, the absolute binding energies
of these levels cannot be guaranteed to be correct for $^{40,41}$K.
This has implications for the relative polarizabilities of the
different isotopes, but has little impact upon the determination
of the polarizability differences of the hyperfine levels
within the same isotope.

\begin{table}
\caption{\label{tab4} The scalar, $\alpha_1$, and tensor, $\alpha^T_1$, dipole
polarizabilities of the hyperfine states of $^{39,40,41}$K. The notation $a[b]$
means $a \times 10^{b}$.  The absolute precision of the polarizabilities
is about 1$\%$ but polarizability differences of different hyperfine states of
the same isotope should be accurate to better than 10$^{-5}$ a.u.
}
\begin{ruledtabular}
\begin {tabular}{lccc}
State     & $F$ & $\alpha_1$ (a.u.) &   $\alpha^T_1$ (a.u.)     \\
\hline
$^{39}$K  $4s_{1/2}$ &   1   &   290.0493839     &    0.102[$-$5]         \\
$^{39}$K  $4s_{1/2}$ &   2   &   290.0499846     & $-$0.481[$-$5]        \\
\hline
$^{40}$K  $4s_{1/2}$ &   7/2  &  290.0505970     &    $-$0.103[$-$4]      \\
$^{40}$K  $4s_{1/2}$ &   9/2  &  290.0489242     &    $-$0.166[$-$4]    \\
\hline
$^{41}$K $4s_{1/2}$ &    1   &   290.0493810      &    0.668[$-$6]         \\
$^{41}$K  $4s_{1/2}$ &   2   &   290.0497114      & $-$0.242[$-$5]        \\
\end{tabular}
\end{ruledtabular}
\end{table}

\begin{table}
\caption{\label{tab5} The hyperfine Stark shifts
of the ground states of $^{39,40,41}$K.
The notation $a[b]$ means $a \times 10^{b}$. }
\begin{ruledtabular}
\begin {tabular}{lc}
Method   &  $\Delta \alpha_1$ (a.u.)     \\ \hline
   &  \multicolumn{1}{c}{$^{39}$K: \ $\alpha_1(F=2)-\alpha_1[(F=1)$} \\
Present                      & 6.007[$-$4]  \\
MBPT-SD \cite{safronova08b}  & 5.996[$-$4]   \\
Perturbation theory \cite{kelly73a}  & 4.9[$-$4]   \\
Perturbation theory \cite{lee75a}  & 5.49[$-$4]   \\
Experiment \cite{snider66a}  &    6.11(61)[$-$4]   \\
Experiment \cite{mowat72a}  &   5.7(2)[$-$4]   \\ \hline
  &  \multicolumn{1}{c}{$^{40}$K: \ $\alpha_1(F=9/2)-\alpha_1(F=7/2)$} \\
Present                      & 1.673[$-$3]  \\ \hline
 &  \multicolumn{1}{c}{$^{41}$K: \ $\alpha_1(F=2)-\alpha_1(F=1)$} \\
Present                      & 3.305[$-$4]  \\
\end{tabular}
\end{ruledtabular}
\end{table}

\subsection{Dipole polarizability }

The dynamic dipole polarizabilities can be computed with the
usual oscillator strength sum-rules.  This is
\begin{equation}
\alpha_{1}(\omega) = \sum_i \frac{ f^{(1)}_{gi} } {\varepsilon_{gi}^2 - \omega^2} \ .
\label{polar1}
\end{equation}
The sum over $i$ includes all allowable fine-structure and hyperfine-structure
allowed transitions.   The $\omega \to 0$ limit of Eq.~(\ref{polar1}) is the
static dipole polarizability.  The dipole polarizability also has a tensor component
for states with $F > 1/2$.  This can be written
\begin{eqnarray}
\alpha^{\rm T}_{1}(\omega) &=&   6\left( \frac{5F_g(2F_g-1)(2F_g+1)}{6(F_g+1)(2F_g+3)}  \right)^{1/2} \nonumber \\
    & \times & \sum_{i}  (-1)^{F_g+F_i}
  \left\{ \begin{array}{ccc}
   F_g & 1 & F_i \\
   1 & F_g & 2
   \end{array} \right\}
\frac{ f^{(1)}_{gi} } {\varepsilon_{gi}^2 - \omega^2}
\label{polar2}
\end{eqnarray}
Static scalar and tensor dipole polarizabilities for the potassium isotope
ground states are listed in Table \ref{tab4}.
The polarizability for a state with non-zero angular momentum $F_g$ depends on the
magnetic projection $M_g$, these can be calculated using the formula \cite{mitroy10a},
\begin{equation}
\alpha_{1,M_g}(\omega) = \alpha_{1}(\omega) + \alpha^{T}_1(\omega) \frac{3M_g^2-F_g(F_g+1)}{F_g(2F_g-1)},
\label{polar3}
\end{equation}
and numerical values are given in the supplementary data.

Table \ref{tab5} gives the hyperfine Stark shift, i.e. the difference between
the polarizabilities for two states with the same $(L,J)$ but different $F$
quantum numbers.  There have been some previous investigations of the hyperfine
Stark shift for $^{39}$K \cite{snider66a,mowat72a,kelly73a,lee75a,safronova08b}.
The polarizability differences in Table \ref{tab5} are given in a.u.  The
hyperfine Stark shift is often reported experimentally as a Stark shift coefficient $k$,
with units of (Hz/(V/m)$^2$.  This is converted into a.u. by multiplying by
0.4018778$\times 10^{8}$ \cite{mitroy10a}.  The present hyperfine Stark shift, is in very good agreement
with the value from a singles plus doubles all-order relativistic many body perturbation
theory calculation \cite{safronova08b} and about 5$\%$ different than the
experimental values \cite{snider66a,mowat72a}.
Differences with earlier calculations are hard to assess since useful data such
as the resonant oscillator strength were not given \cite{kelly73a,lee75a}.

The changes to the dipole matrix elements were an important
part of the polarizability calculation. The polarizability difference
between the $F=1$ and $F=2$ states of $^{39}$K was 6.007$\times 10^{-4}$
a.u. However, omitting the matrix element correction resulted in a
hyperfine Stark shift of about half this size, namely of 3.306$\times 10^{-4}$ a.u.
The changes in the polarizabilities for
$F=1$ and $F=2$ states of $^{39}$K due to change in the matrix elements
were $-$1.688$\times 10^{-4}$ a.u.  and 1.013$\times 10^{-4}$ a.u., respectively.

The polarizabilities in Table \ref{tab4} need to be interpreted properly.
The overall uncertainty in the polarizability is larger than
the precision to which the values have been reported would imply.  The
uncertainty is about 1$\%$ \cite{mitroy10a,jiang13a,safronova08a}.   However,
the polarizability differences between different hyperfine states are
essentially perturbative calculations despite the fact the current
methodology involves the calculation of two polarizabilities followed
by a subtraction.  So the hyperfine Stark shifts can be given to an accuracy
of at least $10^{-4}$ a.u. even though the overall precision in the calculated
polarizabilities are closer to 1 a.u.

Table \ref{tab5} also gives the hyperfine Stark shifts for the ground states of
$^{40}$K and $^{41}$K.  There does not appear to be any experimental or theoretical
information on the hyperfine Stark shifts for either of these isotopes.  The
hyperfine Stark shift for $^{40}$K is close to three times larger than the hyperfine
Stark shift for $^{39}$K.  This is a consequence of the larger hyperfine interaction
constants of $^{40}$K.  The $^{41}$K isotope has the smallest hyperfine interaction
constants and the smallest Stark shift.

The calculations of the polarizabilities for $^{40,41}$K isotopes take into
account the isotope shifts (IS) of the $4s$, $4p_J$ and $5p_{J}$ states.  The
energy intervals between these levels incorporate the isotope shifts into
the polarizability calculation.  However, matrix elements were not adjusted to take
into account the overall isotope shift.  This means that not much significance can be
attributed to the differences in the polarizabilities of the potassium ground
states for the different isotopes.  These could be estimated if the absolute shift
in the ionization energies of these three isotopes were known.  However, the isotope
shifts of the ground state ionization energies are not known so the isotope shift in
polarizabilities of the ground states cannot be estimated.  But to reiterate,
the present methodology does allow for the difference in polarizabilities
due to hyperfine structure to be estimated correctly.

All of the potassium isotopes have a non-zero nuclear spin, so all of the
ground states will have a tensor polarizability which is listed in
Table \ref{tab4}. The $^{40}$K isotope, with the largest nuclear spin, has
the largest tensor polarizabilities.  But even for this system, the tensor
polarizability does not exceed $2 \times 10^{-5}$ a.u. in magnitude.  The
differences between the polarizabilities for the different magnetic sub-levels
of the hyperfine states did not exceed $2 \times 10^{-5}$ a.u.  Polarizabilities
for the different hyperfine magnetic sub-levels are listed in the supplementary
data.

\begin{figure}[tbh]
\centering{
\includegraphics[width=8.5cm,angle=0]{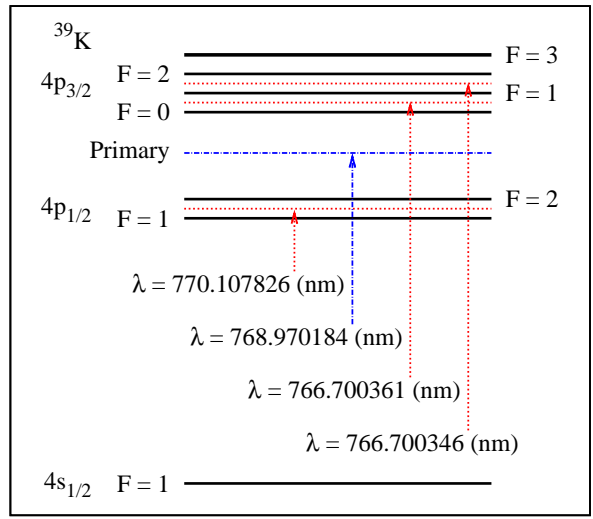}
}
\caption[]{
\label{fig1}
(color online) The energy levels and tune-out wavelengths
for the $4s_{1/2,F=1}$ state of $^{39}$K.  The diagram is not to scale.
The position of the primary tune-out wavelength is indicated.
The tune-out wavelengths for the $4s_{1/2,F=2}$ state are very
similar, but there is no tune-out wavelength between the $F = 0$ and
$F = 1$ states of $4p_{3/2}$ while a new tune-out wavelength occurs
between the $F = 2$ and $F = 3$ states.
}
\end{figure}

\subsection{Tune-out wavelengths }

\subsubsection{$^{39}$K }

The tune out wavelengths for the two hyperfine levels of the $4s_{1/2}$
ground state of $^{39}$K are listed in Table \ref{tab6}.   Hyperfine splitting
leads to two new features in the tune-out spectrum.  First of all,
the splitting of the $4s_{1/2}$ ground state into two hyperfine
levels has resulted in two duplicate sets of tune-out wave lengths,
one for the $4s_{1/2,F = 1}$ state and one for the $4s_{1/2,F = 2}$
state.  The tune-out wavelengths are given to
eight digits after the decimal point to ensure all the differences
between all hyperfine generated tune-out wavelengths are given to
at least two digits.

The hyperfine splitting of the $4p_{J,F}$ levels has also resulted in
the creation of additional tune-out wavelengths that arise when the
polarizability contributions from two adjacent hyperfine levels act to cancel
each other.  Figure \ref{fig1} is a schematic energy level diagram
showing the location of all the tune-out wavelengths for the
$^{39}$K $4s_{1/2,1}$ state.  The hyperfine splitting
of the $4p_{1/2,F}$ states has resulted in one additional tune-out
wavelength, located between
the excitation thresholds of the $4p_{1/2,1}$ and $4p_{1/2,2}$
states.  The hyperfine structure associated with the $4p_{3/2}$ state results
in two additional tune-out wavelengths located between the three $4p_{3/2,F}$
levels that can undergo a dipole transition  with the $4s_{1/2,F'}$
hyperfine level.

\begin{table}
\caption{\label{tab6} Tune-out wavelengths (in nm) for the
$4s_{1/2,F=1}$ and $4s_{1/2,F=2}$ states of $^{39}$K.
Calculations are done with $\alpha_1$ defined by Eq.~(\ref{polar1}).
The tune-out wavelengths omitting consideration of hyperfine splitting
are given for comparison in boldface.  Tune-out wavelengths are
given to eight digits after the decimal point to ensure differences
between adjacent wavelengths near the $4p_{3/2,F'}$ states
are given to at least two digits.  The total uncertainty in the
tune-out wavelengths is about 0.001-0.002 nm as discussed in
the text.}
\begin{ruledtabular}
\begin {tabular}{cc}
 \multicolumn{1}{c}{$4s_{1/2,F=1}$ }   &    \multicolumn{1}{c}{$4s_{1/2,F=2}$ }   \\
\hline
   {\bf 768.97076090    }  &   {\bf 768.97076090    }   \\
  770.10782637     &   770.10870313       \\
   768.97018480  &   768.97110657     \\
   766.70036069     &   766.70125931       \\
   766.70034560     &   766.70123125     \\
   {\bf 405.91731438   } &    {\bf 405.91731438 }  \\
   405.91716336  &   405.91740499    \\
   404.83548479  &   404.83573392        \\
   {\bf 404.72176294} &    {\bf 404.72176294}  \\
   404.72160507   &  404.72185767       \\
   404.52831645     &   404.52856786    \\
   404.52831507  &  404.52856533     \\
\end{tabular}
\end{ruledtabular}
\end{table}

There is one tune-out wavelength for any transition which we define as the
primary tune-out wavelength.  This wavelength is the one which most closely
corresponds to the tune-out wavelengths calculated without hyperfine
splitting.  For the $F = 1$ state of $^{39}$K it is indicated on Figure
\ref{fig1} as the wavelength
of 768.970185  nm  which lies between the excitation thresholds
of the $4p_{1/2}$ and $4p_{3/2}$ states.   The primary tune-out wavelength
for the $4s_{1/2,F = 2}$ state of $^{39}$K is 768.971107 nm.
The absolute precision of these two tune-out wavelength estimates should be
about 0.001-0.002 nm \cite{jiang13a}.  However, the differences between
these two wavelengths is known to a much higher precision.
The difference in the energies for these two tune-out wavelengths is very
closely related to the hyperfine splitting between these two levels.
Table \ref{tab9} compares the energy differences between the two
tune-out primary wavelengths and the hyperfine energy splitting of the
two $4s_{1/2}$ hyperfine states.  They are in agreement to better than
$1.5\%$ and this implies increased precision in the calculated hyperfine
shifts of the tune-out frequencies.

While the calculations of the hyperfine Stark shift are critically reliant
on the use of the energy adjusted reduced dipole matrix elements, this is not true
for the tune-out wavelengths.   For example, the adjustments to the reduced
matrix element made a contribution of $1 \times 10^{-6}$ nm  to the
$^{39}$K $F = 1$ tune-out wavelength of 768.970185  nm.
The change in the matrix element contributed $8 \times 10^{-6}$ nm to
the 405.917163  nm tune-out wavelength.

This information above suggests a very simple approach to determine the hyperfine
splitting of the primary tune-out wavelengths that are located between the two states
of the spin-orbit doublet of the first excited state.  The frequency
difference would just be the energy difference in the $4s_{1/2}$ hyperfine ground
states, i.e.
\begin{equation}
\delta \omega_{\rm tune-out} \approx \delta E_{\rm hfs}(4s_{1/2}) \ .
\end{equation}
The difference between the positions of the primary tune-out wavelengths with
and without hyperfine splitting is also almost equal to the hyperfine energy
shifts.  For example, the tune-out frequency omitting hyperfine splitting is
0.0592523862 a.u.  The tune-out frequency for the $F = 1$ state is
0.0592524306 a.u.  The energy difference of 4.44$\times 10^{-8}$ a.u. is almost
the same as the hyperfine energy shift of 4.39$\times 10^{-8}$ a.u.

The tune-out wavelengths also depend on the magnetic
sub-level of potassium that is occupied.   The tune-out wavelengths associated
with the different magnetic sub-levels of the $4s_{1/2,F}$ states are listed
in the supplementary data.  The tune-out wavelengths for the  $M_F = 0$
and $M_F = 1$ states of $^{39}$K($4s_{1/2,F=1})$ were 768.9701808 nm
and 768.9701868 nm respectively.  The differences in the tune-out wavelengths
for any of the different magnetic sub-levels do not exceed $2 \times 10^{-5}$ nm.

\subsubsection{$^{40,41}$K }

All properties involving $^{40,41}$K should be interpreted with the contents
of the previous section in mind.  First of all, the hyperfine Stark shift
calculations for these isotopes can expected to have an overall accuracy
similar to those of $^{39}$K.  However, not much significance can attributed
to the differences in the polarizabilities of the $^{39,40,41}$K ground states.

The tune-out wavelengths for $^{40}$K and $^{41}$K are given in Table \ref{tab7}.
The tune-out
wavelengths without any hyperfine splitting are also tabulated.  The differences
between the tune-out wavelengths for the different hyperfine states is largest
for $^{40}$K.  This is expected since $^{40}$K has the largest hyperfine
interaction constants.

\begin{table}
\caption{\label{tab7} Tune-out wavelengths (in nm) for the
$4s_{1/2,F=7/2}$ and $4s_{1/2,F=9/2}$ states of $^{40,41}$K.
Calculations are done with $\alpha_1$ defined by Eq.~(\ref{polar1}).
The tune-out wavelengths omitting consideration of hyperfine splitting
are given for comparison in boldface.   }
\begin{ruledtabular}
\begin {tabular}{ccc}
 \multicolumn{1}{c}{$^{40}\text{K}(4s_{1/2,F=7/2})$ } &  &    \multicolumn{1}{c}{$^{40}\text{K}(4s_{1/2,F=9/2})$ }   \\
  \cline{1-1}  \cline{3-3}
   {\bf 768.97051283} & &   {\bf 768.97051283}  \\
    770.10942041 &    &   770.10697923\\
    768.97193896 &   &    768.96937193    \\
    766.70202758  &   &    766.69955766    \\
    766.70195723  &  &  766.69947243 \\
   {\bf 405.91718563} &  &  {\bf 405.91718563} \\
    405.91755943 &    &   405.91688660 \\
    404.83589207 &    &  404.83519815 \\
   {\bf 404.72163455 } &   &  {\bf 404.72163455 }  \\
     404.72202530 &    &    404.72132195 \\
     404.52873087 &    &    404.52803363  \\
     404.52872456  &   &  404.52802596 \\ \hline
 \multicolumn{1}{c}{$^{41}\text{K}(4s_{1/2,F=1})$ } &  &    \multicolumn{1}{c}{$^{41}\text{K}(4s_{1/2,F=2})$ }   \\
  \cline{1-1}  \cline{3-3}
      {\bf 768.97029599 } &  &  {\bf 768.97029599}  \\
    770.10759459  &   &    770.10807699 \\
    768.96997907 &    &    768.97048613 \\
    766.70013234 &    &    766.70062962     \\
     766.70012847  &   &   766.70061632  \\
   {\bf 405.91706559 } &    &    {\bf 405.91706559} \\
    405.91698251 &     &   405.91711543 \\
    404.83530536     &   &    404.83544246 \\
  {\bf 404.72151479}  &  &  {\bf 404.72151479}  \\
     404.72142795   &   &  404.72156690 \\
     404.52813758   &   &  404.52827616  \\
     404.52813722   &   &  404.52827496 \\
\end{tabular}
\end{ruledtabular}
\end{table}

The shifts in energies have a much larger effect on the tune-out wavelengths
than the changes in reduced matrix elements.  So differences between the
tune-out wavelengths for different isotopes can be estimated from the data
in Tables \ref{tab6} and \ref{tab7} to a precision of about
10$^{-5}$ nm.

\begin{table}
\caption{\label{tab9}
Comparison of the $4s_{1/2}$ state hyperfine energy splittings and the
energy splitting of the primary tune-out frequencies, $\omega_{\rm to}$
mainly due to hyperfine splitting of the $4s_{1/2}$ level.
The notation $a[b]$ means $a \times 10^{b}$. }
\begin{ruledtabular}
\begin {tabular}{lcc}
Transition  & $\Delta E_{\rm hfs}$ (a.u.)  &  $\Delta \omega_{\rm to}$ (a.u.) \\ \hline
$^{39}$K $W_{F=2}- W_{F=1}  $ & 7.017[$-$8]  &   7.103[$-$8]\\
$^{40}$K $W_{F=7/2}- W_{F=9/2} $ & 1.954[$-$7]  &  1.978[$-$7]\\
$^{41}$K $W_{F=2}- W_{F=1}  $ & 3.861[$-$8]  &  3.907[$-$8]\\
\end{tabular}
\end{ruledtabular}
\end{table}

\section{Conclusions}

The impact of hyperfine structure on the tune-out wavelengths of three
isotopes of potassium have been calculated.  The hyperfine structure of the
ground and excited states leads to a number of additional tune-out wavelengths.
The additional tune-out frequencies associated with the hyperfine splitting
of the $np_J$ excited states would be difficult to detect due to the very small
energy splittings of the hyperfine levels.  The additional tune-out wavelengths
associated with the splitting of the $4s$ ground state are more likely to be
detectable in an experiment.  The
splittings in the tune-out wavelengths for the primary tune-out wavelength of
$^{39,40,41}$K are 0.0008 nm, 0.0027 nm and 0.0005 nm respectively.  The
uncertainty in a recent experiment was 0.0015 nm \cite{holmgren12a}.  The
hyperfine splitting in the tune-out wavelengths would become apparent
with a modest increase in experimental precision.

The method used to determine the shifts in the tune-out frequency was
unorthodox, being essentially a second order calculation using energy
and matrix element shifts applied prior to the evaluation of the
oscillator strength sum rules.  This simplified calculation was adequate
for a system with $(L,J) = (0,1/2)$ quantum numbers and a nucleus with
a small quadrupole moment and the agreement with a previous all-order MBPT
calculation \cite{safronova08b} could hardly have been better.  The method
could be made more rigorous without a great deal of effort.  It should be
possible to incorporate the hyperfine operator into the Hamiltonian prior
to diagonalization.  The additional computational efforts would not be
prohibitive.

One result of the present analysis is that the tune-out wavelength is
largely insensitive to the small changes in the matrix elements resulting
from the hyperfine interaction.  The hyperfine interaction driven changes
in the tune-out wavelengths are most sensitive to the changes in the
transition frequencies.   Indeed, the dominant effect on the tune-out
wavelengths can probably be estimated by just taking into account
the energy splits and shifts caused by the hyperfine interaction.

This research was supported by the Australian Research Council
Discovery Project DP-1092620.  We thank Dr Marianna Safronova for
suggesting we investigate the shifts in the tune-out wavelengths caused
by the hyperfine interaction.


\end{document}